\documentclass[fleqn,usenatbib]{mnras}

\usepackage{newtxtext,newtxmath}

\usepackage[T1]{fontenc}

\DeclareRobustCommand{\VAN}[3]{#2}
\let\VANthebibliography\thebibliography
\def\thebibliography{\DeclareRobustCommand{\VAN}[3]{##3}\VANthebibliography}

\usepackage{ulem}
\usepackage{amsmath}  
\usepackage{amsfonts} 
\usepackage{graphicx} 
\usepackage{pgfplots}
\usepackage{pgfplotstable}
\usepackage{booktabs}
\usepackage{array}
\usepackage{colortbl}
\usepackage{graphicx}
\usepackage{caption}
\usepackage{graphicx}
\usepackage{dcolumn}
\usepackage{bm}
\pgfplotsset{
  width = 12cm, compat=1.9
}

\title[Virial theorem for highly magnetized stars]{Modified virial theorem for highly magnetized white dwarfs}

\author[Mukhopadhyay, Sarkar \& Tout]{
Banibrata Mukhopadhyay,$^{1}$\thanks{E-mail: bm@iisc.ac.in}
Arnab Sarkar$^{2}$
and Christopher A. Tout$^{3}$
\\
$^{1}$Department of Physics, Indian Institute of Science, Bangalore  560012, India\\
$^{2}$Department of Physical Sciences, Indian Institute of Science Education and Research, Kolkata 741246, India\\
$^{3}$Institute of Astronomy, The Observatories, Medingley Road, Cambridge CB3 OHA, UK
}

\date{Accepted XXX. Received YYY; in original form ZZZ}

\begin{document}
\label{firstpage}
\pagerange{\pageref{firstpage}--\pageref{lastpage}}
\maketitle








\begin{abstract}

Generally the virial theorem provides a relation between various
components of energy integrated over a system.  This helps us to
understand the underlying equilibrium.  Based on the virial theorem we
can estimate, for example, the maximum allowed magnetic field in a
star.  Recent studies have proposed the existence of highly magnetized
white dwarfs, with masses significantly higher than the Chandrasekhar
limit.  Surface magnetic fields of such white dwarfs could be more
than $10^{9}\,$G with the central magnitude several orders higher.
These white dwarfs could be significantly smaller in size than their
ordinary counterparts (with surface fields restricted to about
$10^9\,$G).  In this paper we reformulate the virial
theorem for non-rotating, highly magnetized white dwarfs (B-WDs) in
which, unlike in previous formulations, the contribution of the
magnetic pressure to the magnetohydrostatic balance cannot be
neglected.  Along with the new equation of magnetohydrostatic
equilibrium, we approach the problem by invoking magnetic flux
conservation and by varying the internal magnetic field with the
matter density as a power law.  Either of these choices 
are supported by previous independent work and neither violates any
important physics.  They are useful while there is no prior knowledge of field profile
within a white dwarf.  We then compute the modified gravitational,
thermal and magnetic energies and examine how the magnetic pressure
influences the properties of such white dwarfs.  Based on our results
	we predict important properties of these B-WDs, which 
	turn out to be independent of our chosen field profiles.

\end{abstract}


\maketitle 

\section{Introduction} 

White dwarfs are electron degenerate compact stars in which the
outward degeneracy pressure force is able to balance the inward
gravitational force only when the white dwarf mass is below the
Chandrasekhar limit \citep{Chandrasekhar1935}.  In Newtonian
calculations, the limiting mass of a carbon--oxygen white dwarf is
$1.44\,M_\odot$, where $M_\odot$ is the mass of Sun, but this can be
increased by rotation or magnetic fields \citep{Ostriker1968}.  White
dwarfs are considered to be the progenitors of the type~Ia supernovae
which are some of the most widely studied astronomical events and
explosions and rightfully so because of their usefulness to measure
cosmic distances.  Although not everything is understood about these
events, the general consensus is that they are thermonuclear
explosions of white dwarfs with masses very close to the Chandrasekhar
limit.  However recent observations of a fast-growing number of
several peculiar, overluminous type~Ia supernovae, for example
SN~2006gz, SN~2007if, SN~2009dc and SN~2003fg, bring even this basic
idea into serious question, because they are best explained by
progenitor white dwarfs with super-Chandrasekhar masses in the range
$2.1$ to~$2.8\,M_\odot$ \citep{AndrewHowell2006,Scalzo2010}.

One acceptable proposal is that these super-Chandrasekhar white dwarfs
are highly magnetized white dwarfs (B-WDs) \citep{Das2013}.  This
model was first formulated by Mukhopadhyay and his collaborators
\citep{KUNDU2012,DAS2012}, who put the idea in the limelight.  It has
since been elaborated upon by various groups.  Although the idea was
further questioned by various authors, most, if not all, of the
concerns that were raised against B-WDs have been addressed by
Mukhopadhyay and his collaborators in subsequent publications.  For
instance, it was shown by \citet{Das2014, Das2014_1} that the
unreasonable possibility of a $24\,M_\odot$ white dwarf
\citep{Dong2014, Coelho2014} is ruled out when magnetic energy density
is appropriately included in magnetostatic balance and the mass
equation simultaneously and self-consistently.  Also, the maximum
possible magnetic field sustainable in a B-WD and the energy content
of various corresponding terms in the virial theorem
\citep{Coelho2014} were shown \citep{Das2014} to be misleading, unless
the virial theorem is established for a strong field.  This was
briefly explored by \citet{mukhopadhyay2016soft} and we elaborate upon
it here in detail.  Other authors \citep{Chatterjee2017} argued, based
on hypothetical pycnonuclear reaction rates, that super-Chandrasekhar
$\!^{16}$O~B-WDs are not possible but instead limited to less than
$1.38\,M_\odot$.  However, if their chosen pycnonuclear reaction rates
are correct, even the Chandrasekhar limit for nonmagnetic,
non-rotating white dwarfs has to be restricted to $1.21\,M_\odot$.
This is counter-intuitive.  However the pycnonuclear reaction rates
are extremely uncertain and must be constrained carefully, more so
than they have been.  Moreover with the same pycnonuclear reaction
rates $\!^{12}$C~white dwarfs recover the Chandrasekhar limit and the
underlying B-WDs are found to be super-Chandrasekhar.  In fact later
on, with more logical pycnonuclear reaction rates, \citet{Otoniel2019}
showed that highly magnetized non-rotating super-Chandrasekhar white
dwarfs are quite possible with masses grater than $2\,M_\odot$.  So
the issue of the pycnonuclear reaction is no longer problematic.  On
the other hand, it has been well-known since 1973 \citep{Tayler1973,
  Markey1973} that purely poloidal (or poloidally dominated) or purely
toroidal fields are unstable.  Hence any stability analysis of a
magnetized star based on purely poloidal and purely toroidal field, as
attempted by \citet{Bera2016}, does not make any sense
\citep{mukhopadhyay2016significantly}.  However, it was shown by
\citet{Wickramasinghe2013} that a white dwarf with toroidally
dominated mixed field configuration (with a small poloidal component)
remains stable for a long time.  In this case the white dwarf is
approximately spherical in shape \citep{Subramanian2015}.  Hence, in a
convenient stable mixed field configuration, we can always formulate a
profile for the magnitude of magnetic field with respect to the
stellar radius or density as we advocate here.  Without prior
knowledge of the variation of field magnitude within a star our
simpler profiles chosen below are not at odds with any known physics
or observations.

Nevertheless, there is another extensive set of independent work that
supports the proposition of super-Chandrasekhar magnetized white
dwarfs \citep[][to list a significant selection]{Federbush2014,
  Franzon2015, Sotani2017, Franzon2017, Shah2017, Moussa2017,
  Roy2019}.  Recently, the Mukhopadhyay group explored the luminosity
and possible gravitational radiation of rotating B-WDs.  While
\citet{Bhattacharya2018} and \citet{Gupta2020} showed that the B-WDs
turn out to be too dim to detect, \citet{kalita2019continuous}
explored their gravitational waves, along with electromagnetic
counterparts, of which the detectability by forthcoming instruments was further
explored by \citet{kalita2019continuous} and \citet{Kalita2020}.
Moreover, others often go beyond the idea of introducing strong
magnetic fields and invoke additional physics, such as anisotropic
pressure \citep{Herrera2013}, lepton number violation
\citep{Belyaev2015}, modified gravity \citep{Banerjee2017,EslamPanah2019},
effects of net charge \citep{Liu2014,Carvalho2018} and ungravity
\citep{Bertolami2016}.

Here we explore the maximum possible magnetic fields and the mass
needed to sustain a magnetized star, particularly a white dwarf, by
modifying the virial theorem.  Note that strong magnetic fields are
expected to affect not only momentum balance (by magnetostatic balance
for example) but also the underlying equation of state and thermal
energy.

The virial theorem relates the integrated gravitational potential,
kinetic, thermal and magnetic energies and provides an insight into
the equilibrium of the system.  Understanding the virial theorem for
B-WDs helps us understand how high a magnetic field could be sustained
therein and plausibly modify the properties of a normal white dwarf,
thereby, making it a B-WD.  In the weak field regime, the deviation of
the Chandrasekhar limit owing to magnetic field has been explored
earlier by \citet{Shapiro:1983du} and any changes to the properties of
the white dwarf found to be only perturbative.

In the next section we recall the basic idea of scalar virial theorem,
assuming the magnetic field is not perturbative.  Subsequently, based
on the magnetohydrostatic (rather than the hydrostatic) equilibrium,
we compute various energy terms in virial equilibrium in
section~\ref{secmodels} for two model magnetic field profiles. 
In the beginning of the same section we also justify the chosen 
field profiles.  In
section~\ref{secresults} we discuss the results in detail and we end
with conclusions in section~\ref{secconclusions}.

\section{The scalar virial theorem}
\label{secvirial}

The virial theorem is a general integral theorem which relates various
components of energy.  We use it to discuss the effects of high
magnetic fields on white dwarfs, thereby making them B-WDs. The well known
form of the virial theorem can be recalled as
\begin{center}
\begin{equation} 2T + W + 3 \Pi + \mu = 0 \end{equation}
\end{center}
for a dynamically stable star when the moment of inertia $I$ is constant
(see \citealt{Shapiro:1983du} or \citealt{Eldridge2019} for details).
Here,
\begin{center}
\begin{equation}
\label{I}
I = \displaystyle{\frac{1}{2} \int_V \rho\,x^2d^3x} \end{equation}
\end{center}
is a generalized moment of inertia and
\begin{center}
\begin{equation} T = \displaystyle{\frac{1}{2} \int_V \rho\,v^2d^3x}, \end{equation}
\end{center}
\begin{center}
\begin{equation} \Pi = \displaystyle{\int_V P\,d^3x}, \end{equation}
\end{center}
\begin{center}
\begin{equation}
\label{mu}
\mu = \displaystyle{\frac{1}{8\pi} \int_V B^2\,d^3x} \end{equation}
\end{center}
and
\begin{center}
\begin{equation}
\label{W1}
W = -\displaystyle{\int_M \frac{Gm\,dm}{r}}, \end{equation}
\end{center}
are the kinetic, thermal, magnetic and gravitational energies
respectively, when $\rho$ is the density, $v$ is the bulk velocity,
$P$ is the pressure of stellar matter, $B$ is the magnetic field, $G$
is Newton's gravitational constant, $M$ is the mass of the star of volume
$V$, $r$ is the radius from the centre of the star, $dm$ is the
elemental mass and $d^3x$ is the corresponding volume.

Here we consider the case of a (static) non-rotating white dwarf for
which $T=0$. Hence, the scalar virial theorem reduces to
\begin{center}
\begin{equation}
\label{virial1}
W + 3\Pi + \mu = 0, \end{equation}
\end{center}
and this can be approximately recast to
\begin{center}
\begin{equation}
\label{virial2}
-\alpha \displaystyle{\frac{GM^2}{R}} + \beta^\prime M
\displaystyle{\frac{P}{\rho}} + \gamma
\displaystyle{\frac{\Phi_M^2}{R}} = 0
\end{equation}
\end{center}
\citep{Shapiro:1983du,mukhopadhyay2016soft}, where $\alpha$,
$\beta^\prime$ and $\gamma$ are the constants, determined by the shape
and other properties of the star investigated below, the magnetic flux 
through its surface ${{\Phi}_M} \approx \overline{B}R^2$, with $\overline{B}$ 
being the average magnetic field and  $R$ is the radius
of the star.  Here we consider the isotropic effects of an averaged
magnetic field $B$ and so overall consider the star to be spherical in
shape.  For the plausibility of this, see numerical simulation results
by \citet{Wickramasinghe2013} and \citet{Subramanian2015},
particularly for toroidally dominated cases.  

Next we assume that a polytropic equation of state (EoS) is satisfied
through the entire star such that $P = K {\rho}^{\Gamma}$, where $K$
and $\Gamma$ are the polytropic constants and $M = {\frac{4}{3} \pi
  R^3 \rho}$, where $\rho$ is the mean
density.  The scalar virial theorem can then be reduced to
\begin{center}
\begin{equation}
\label{virial3}
- \alpha \displaystyle{\frac{GM^2}{R}} + \beta
\displaystyle{\frac{M^{\Gamma}}{R^{3{(\Gamma - 1)}}}} + \gamma
\displaystyle{\frac{\Phi_M^2}{R}} = 0,
\end{equation}
\end{center}
where $\beta = K(3/4\pi)^{\Gamma - 1}\beta'$.  We have simply substituted $P$ from the EoS in
equation~(\ref{virial2}) to arrive at the second term in equation~(\ref{virial3}).

Now rearranging equation~(\ref{virial3}), we obtain
\begin{center}
\begin{equation}
\label{M1}
M = \sqrt{\displaystyle{\frac{ \gamma \Phi_M^2}{\alpha G\Big(1 -
      \displaystyle{\frac{\beta M^{\Gamma - 2}}{\alpha GR^{3\Gamma -
            4}}\Big)}}}},
\end{equation}
\end{center}
for any $\Gamma$.  For $\Gamma = 4/3$, appropriate for extremely
relativistic degenerate electrons, this becomes
\begin{center}
\begin{equation}
M = \sqrt{\displaystyle{\frac{ \gamma \Phi_M^2}{\alpha G\Big(1 -
      \displaystyle{\frac{\beta M^{-2/3}}{\alpha G}\Big)}}}} ,
\end{equation}
\end{center}
which is independent of $R$ for a fixed magnetic flux, as expected
from Chandrasekhar's theory.  This can be solved for $M$.  For
$\Gamma=2$, appropriate to high magnetic field and high density
\citep{Das2013}, equation~(\ref{M1}) becomes
\begin{center}
  \begin{equation}
    M = \sqrt{\displaystyle{\frac{ \gamma \Phi_M^2}{\alpha G\Big(1 -
          \displaystyle{\frac{\beta}{\alpha GR^{2}}\Big)}}}},
\end{equation}
\end{center}
to give the mass explicitly.


\section{Modification to the virial theorem }
\label{secmodels}

Here we evaluate the coefficients $\alpha$, $\beta$ and $\gamma$ to
establish the virial theorem at high magnetic field.  First we note,
very importantly, that in the presence of strong magnetic field, the
upper limit for magnetic fields in white dwarfs, as discussed for weak
field cases by \citet{Shapiro:1983du}, must be revised because the contribution of
the magnetic pressure to the magnetohydrostatic balance equation
cannot be neglected.  Here we attempt to revise it in a simpler
framework.  The new momentum balance condition, neglecting the effect
of magnetic tension, is given by \citep[see for example][]{Das2014}
\begin{center}
\begin{equation}
\label{mhd_eq}
\displaystyle{\frac{1}{\rho}\displaystyle{\Big(\frac{dP}{dr} +
    \frac{dP_B}{dr} \Big)} = -\frac{Gm(r)}{r^2}}
 \end{equation}
\end{center}
at an arbitrary radius $r$ with mass enclosed at that radius $m(r)$,
where $\rho$ includes the contribution from magnetic field and
  $P_B$ is the pressure owing to the magnetic field of the star.
Neglect of the magnetic tension for now is justified because our
interest is to estimate the maximum possible magnetic field strength
and its effect in white dwarfs, without worrying about underlying
stability issues or the shape of the star. Indeed this implies
  neglecting anisotropic effects which would further make the star
  nonspherical in a manner that cannot be addressed with a scalar
  virial theorem. Note that terms associated with magnetic pressure
and magnetic tension are of the same order of magnitude and the virial
theorem deals with the effects of order of magnitude by its
virtue. 

We use two different approaches to address this problem based on
equation~(\ref{mhd_eq}): first
we invoke flux conservation (freezing) which is quite common in stars
when conductivity is high and secondly we assume $B$ to vary as a
power law with respect to density, just as the EoS of thermal
pressure, throughout.  This choice, without other prior knowledge of the field
profile within the star, does not violate any important
physics, e.g. no magnetic monopoles or spherical magnetohydrostatic 
equilibrium, while indeed magnetic field is expected to be related to the
matter density. Below we justify the choice of these two approaches
and the underlying field profiles.

\subsection{Physical justification of field profiles\label{secfieldjust}}

While the surface field of a star can be observationally inferred or
even determined, there is no reliable practice to infer its interior
field.  However there is ample evidence that stars exhibit dipolar
field geometries, at least in their outer regions.  Therefore such
stars are expected to have stronger interior fields than at the
surface, following a power law with respect to the radial coordinate.
See, e.g., \citealt{fendt,pili,das15,tout,pons,Otoniel2019}, for a few
representative examples in neutron stars and white dwarfs.  Thus the
field magnitude in a white dwarf could certainly follow a scaling as
$B\propto r^{-m}$, with $m=3$ corresponding to dipole.  For $m=3$, $B$
effectively scales as the inverse square of the stellar size because
the magnetic moment is proportional to the size of the star.  In
general, white dwarfs and all stars are expected to exhibit much more
complicated multipolar geometries combining poloidal and toroidal
field components.  Numerical simulations show that the central field
of a white dwarf could be several orders of magnitude higher than the
surface field (\citealt{Subramanian2015,mukhopadhyaySco,tout}).  In
fact, recently \cite{tout} modelled the evolution of two components of
the magnetic field along with angular momentum based on Cambridge
stellar evolution code using three time-dependent advection-diffusion
equations coupled to the structural and compositional equations of
stars.  They found that the magnetic field could be dipolar,
decaying with an inverse square law, in most of the star.  This gives us
greater confidence to choose a model field that
decays with the radial coordinate from the centre following a power
law.  The computations of \cite{tout} also showed that, even late stages of
stellar evolution, large-scale magnetic fields are sustained in degenerate cores and, based
on conservation of magnetic flux, very high fields can develop in white
dwarfs.  Hence, the force owing to magnetic pressure must be considered
in the magnetohydrostatic balance equation~(\ref{mhd_eq}) to correctly
establish the virial theorem of highly magnetized white dwarfs.

Now for the conservation of magnetic flux throughout the star, which
is likely in highly conducting white dwarfs with very thin envelopes,
$Br^2$ is conserved.  This leads to the scaling $B\propto r^{-2}$,
which is quite synonymous to the dipole consideration above.  However
this cannot be strictly valid to the centre of the star because the
field strength cannot be singular there.  Therefore we invoke such a
radial variation of field from the surface to a finite distance inside
the star below which field is assumed to be constant.

Now the density of a star generally decreases with increasing radial
coordinate.  From a simple self-similar consideration the scaling of
density is given by $\rho\propto r^{-3/2}$ (\citealt{narayanyi}).
Therefore, from the above discussion, the magnitude of the magnetic
field should scale with density as $B\propto\rho^p$ with $p>0$.  Hence
it is justified that the magnetic pressure directly scales with the
density, with a similar relation to that of stellar matter pressure.
This argues in favour of our second choice of magnetic field profile
as a power law with matter density.  Indeed the field magnitude within
neutron stars and white dwarfs has been extensively modelled with a
more complex variation with matter density, rather than a simple power
law and this has successfully explained some observations (e.g.
\citealt{debades,Gupta2020}).  Nevertheless such a field profile turns
out to reveal a constant field in the high density regime and one that
decays outside it.  As a white dwarf and generally a star is expected
have a high density core and a low density envelope, this profile
practically mimics the profile described above primarily based on
magnetic flux conservation.

We show here that two apparently different magnetic field profiles,
prescribed based on apparently different physics, give rise to very
similar results.  Hence, the effect of magnetic pressure and corresponding
gradient in the magnetohydrostatic balance in the virial theorem is
independent of the chosen field profile.  Although the calculations described
below rely on the chosen field profiles, and also the chosen EoS, it
appears that our conclusions do not depend on them as long as they are
prescribed based on realistic physics.  Our approach is similar to
invoking an EoS, as commonly done when working with the virial
theorem.

\subsection{Invoking Magnetic Flux Conservation}

First we consider a case of an approximately constant field in the
central region ($B_{\rm int}$), which further falls off from the centre 
towards the surface, as described in section~\ref{secfieldjust}. Further we consider
the central region to be confined to a radius of $R/n$ with 
the field falling off as $r^2$ towards the surface outside.  We apply
flux conservation from $R/n$ to $R$ to calculate the dependence of
$P_B$ on the radius and obtain $B(r)$ as a piecewise smooth
function such that
\begin{equation}
\label{B}
       B(r)=
        \left\{ \begin{array}{ll}
            \displaystyle{\frac{ {{\Phi}_M} n^2}{R^2}} &0\leq r \leq R/n \\
\\
           \displaystyle{\frac{ {{\Phi}_M} }{r^2}} & R/n < r \leq R .
        \end{array} \right.
\end{equation}
It is apparent that the larger $n$ is so the smaller stellar core.
Because $P_B ={B^2}/{8\pi}$, we obtain $P_B$ in terms of $r$ as
\begin{equation}
\label{P_B}
       P_B(r)=
        \left\{ \begin{array}{ll}
            \displaystyle{\frac{ \Phi_M^2 n^4}{8\pi R^4}} &0\leq r \leq R/n \\
\\
           \displaystyle{\frac{ \Phi_M^2 }{8\pi r^4}} & R/n < r \leq R .
        \end{array} \right.
\end{equation}
Thus both $B$ and $P_B$ are continuous functions that are constant in
a central stellar core.  The gravitational energy
\begin{center}
\begin{equation}
\begin{aligned}
\label{W2}
W ={} &- \int_{0}^{R} \displaystyle{\frac{Gm(r)}{r}4\pi r^2\rho\,dr}\\
      &= \int_{0}^{R} 4\pi r^3\,(dP + dP_B)
\end{aligned}
 \end{equation}
\end{center}
and, with $P_B$ given by equation~(\ref{P_B}), we obtain
\begin{center}
\begin{equation}
\begin{aligned}
\label{W3}
W={}&\int_{0}^{R} 4\pi r^3 (dP + dP_B) = [\displaystyle{4\pi r^3 (P +
    P_B)}]_0^R\\ &- 3\int_0^R\displaystyle{4\pi r^2P\,dr} -
3\int_0^R\displaystyle{4\pi r^2P_B\,dr}.
\end{aligned}
\end{equation}
\end{center}
The second term in RHS of equation~(\ref{W3}) is
\begin{center}
\begin{equation}
\begin{aligned}
\label{P_decomp}
{}&-3\int_0^R P4\pi r^2 dr = -3\int_0^R \frac{P}{\rho}\,dm\\ &=
3\left[\frac{P}{\rho}m\right]_0^R + 3\int_0^R\,d\left(\frac{P}{\rho}\right)m.
\end{aligned}
\end{equation}
\end{center}
Because $P/\rho = K\rho^{\Gamma-1}$,
%
\begin{center}
\begin{equation}
\begin{aligned}
\label{imp1}
d\left(\frac{P}{\rho}\right) = \frac{\Gamma-1}{\Gamma}\,\frac{dP}{\rho}
\end{aligned}
 \end{equation}
\end{center}
and using equation~(\ref{mhd_eq}) 
we obtain
\begin{center}
\begin{equation}
\begin{aligned}
\label{imp2}
\frac{dP }{\rho} = \displaystyle{Gm\, d\left(\frac{1}{r}\right)}-\frac{dP_B}{\rho}.
\end{aligned}
 \end{equation}
\end{center}
So, with equations~(\ref{imp1}) and (\ref{imp2}), the last term of equation~(\ref{P_decomp}) is
\begin{center}
\begin{equation}
\begin{aligned}
\label{imp30}
{}& 3\int_0^R\,d\left(\frac{P}{\rho}\right)m =
3\frac{\Gamma-1}{\Gamma}\int_0^R
m\,\frac{dP}{\rho}\\
&=3\frac{\Gamma-1}{\Gamma}\left(\int_0^R m^2 G
d\left(\frac{1}{r}\right) - \int_0^R m\,\frac{dP_B}{\rho}\right).
\end{aligned}
\end{equation}
\end{center}
Because
\begin{center}
\begin{equation}
\begin{aligned}
\label{imp3}
{}& \int_0^R Gm^2\,d\left(\frac{1}{r}\right) =
\left[\frac{Gm^2}{r}\right]_0^R - 2\int_0^R \frac{Gm\,dm}{r},
\end{aligned}
 \end{equation}
\end{center}
using equations~(\ref{P_decomp}), (\ref{imp30}) and~(\ref{imp3}), from
equation~(\ref{W3}) we obtain
\begin{center}
\begin{equation}
\begin{aligned}
\label{W4}
W ={} &{ [4\pi r^3(P+P_B)]}^R_0 -
3{\displaystyle{\left[\frac{Pm}{\rho}\right]_0^R}} +
\displaystyle{\frac{3(\Gamma-1)}{\Gamma}\frac{GM^2}{R}}\\ &
+\displaystyle{\frac{6(\Gamma-1)}{\Gamma}W}
-3\int_{0}^{R}\displaystyle{\frac{(\Gamma-1)}{\Gamma}\,\frac{m
    dP_B}{\rho}}\\ &-3\int_{0}^{R}4\pi r^2\,P_B dr .
\end{aligned}
\end{equation}
\end{center}
This equation has all the significant terms that may or may not
vanish.  In the first term of RHS of equation~(\ref{W4}), only the
$P_B$ part survives at $r=R$, the surface of the star.  The second
term vanishes on the assumption that $P/\rho$ is negligibly small at
$r = R$, compared to other terms because the density is very small at
the surface and $\Gamma > 1$.  This could easily be verified with the
chosen polytropic EoS.  We see that the presence of the fifth and
sixth terms on the RHS of the expression for $W$ are solely due to the
presence of magnetic pressure in addition to the matter pressure.  In
order to integrate the fifth term, we approximate $\rho(r) =
{m(r)}/{\frac{4}{3}\pi r^3}$. Of course in practice $\rho(r)$
  should be a local density, not that averaged over the region from
  centre to $r$, but for the ease of computation we approximate it so
  and note that, in any case, the virial theorem stands on averaged
  effects.  Below it will be evident that this approximation does not
  influence our main conclusion.
Now, we just use $P_B$ from equation~(\ref{P_B}) for the fifth and
sixth terms in order to obtain an expression in terms of
$\Phi_M^2$.  Finally, putting all these together and writing everything
in terms of $W$, we have
\begin{center}
\begin{equation}
\label{W5}
W =  -\displaystyle{\frac{3(\Gamma-1)}{5\Gamma - 6}\frac{GM^2}{R}} + \displaystyle{\frac{2(n-1)}{5\Gamma - 6}\frac{ \Phi_M^2}{R}}.
 \end{equation}
\end{center}
\cite{tout} showed, in numerical simulations, that, as the
  degenerate core grows in an asymptotic giant branch star, the
  magnetic field may not penetrate to the centre because the centre
  becomes superconducting first and the field cannot diffuse inwards
  easily.  In that case $B(r)=0$ in $0\le r\leq R/n$ and $W$ is
  amended with a contribution $\Phi_M^2 n/2R$.

We find the magnetic energy component
\begin{center}
\begin{equation}
\mu = \frac{1}{8\pi}\int_V B^2 dV = \int_0^R 4\pi r^2 P_B\,dr,
 \end{equation}
\end{center}
can be easily integrated to give
\begin{center}
\begin{equation}
\label{mu1}
\mu = \frac{\Phi_M^2}{R}\left(\frac{4n -3}{6}\right).
 \end{equation}
\end{center}
Substituting equations~(\ref{W5}) and~(\ref{mu1}) in equation~(\ref{virial1}), we obtain
\begin{center}
\begin{equation} 
\begin{aligned}
 -\displaystyle{\frac{3(\Gamma-1)}{5\Gamma - 6}\frac{GM^2}{R}} +
 \frac{3^\Gamma KM^\Gamma}{(4\pi R^3)^{\Gamma-1}} \\ + \left(
 \displaystyle{\frac{2(n - 1)}{5\Gamma - 6}} + \displaystyle{\frac{4n
     -3}{6}}\right)\frac{\Phi_M^2}{R} = 0
\end{aligned}
\end{equation}
\end{center}
and, comparing with equation~(\ref{virial3}), we have
\begin{center}
\begin{equation}
\label{alpha_flfr}
\alpha = \displaystyle{\frac{3(\Gamma - 1)}{5\Gamma - 6}},
 \end{equation}
\end{center}
\begin{center}
\begin{equation}
\label{beta_flfr}
\beta = \displaystyle{\frac{3^{\Gamma}K}{{(4\pi)}^{\Gamma - 1}}}, 
 \end{equation}
\end{center}
\begin{center}
\begin{equation}
\label{gama_flfr}
\gamma = \displaystyle{\frac{2(n - 1)}{5\Gamma - 6}} + \displaystyle{\frac{4n -3}{6}} .
 \end{equation}
\end{center}
Note that there is a change in $\gamma$, which is now significantly
greater than for a weakly magnetized white dwarf for which
$\gamma = 1/6$.  For example, with $n=10$ and $\Gamma = 2$, $\gamma =
32/3$.  Note that when $n=1$ the situation simplifies to that of a
non-magnetized or weakly magnetized white dwarf or a B-WD with
constant $B$ and hence constant $P_B$ throughout.

\subsection{Varying $B$ as a Power Law}

Now we instead assume that the variation of $B$ is a power law 
with density.  Thus, the corresponding magnetic pressure $P_B$ = $K_1
{\rho}^{\Gamma_1}$ with $K_1$ and $\Gamma_1$ constant
\citep{mukhopadhyay2016soft}, as justified in section-3.1.  The gravitational energy for such a
star is
\begin{center}
\begin{equation}
\begin{aligned}
\label{Wp}
W ={}& - \int_{0}^{R} \displaystyle{\frac{Gm(r)}{r}4\pi r^2\rho\,dr}\\
&= \int_{0}^{R} 4\pi r^3\,\displaystyle{\Big(dP + dP_B\Big)}.
  \end{aligned}
 \end{equation}
\end{center}
We integrate~(\ref{Wp}) following the same procedure as we
started with for equation~(\ref{W2}), given by equation~(\ref{W3}).
However here the term $[4\pi r^3(P + P_B)]$ vanishes in
both the limits, because the stellar density vanishes at the
surface.  So we drop it from further calculations.  We can write
the remaining terms of equation~(\ref{W3}) as
\begin{center}
\begin{equation}
\begin{aligned}
-3\int_0^R 4\pi r^2(P+P_B)dr={}& - 3\int_{0}^{R} \frac{P}{\rho}\,dm - 3\int_{0}^{R} \frac{P_B}{\rho}\,dm
  \end{aligned}
 \end{equation}
\end{center}
and using equation~(\ref{imp1}) we obtain
\begin{eqnarray}
\nonumber
\label{W6}
W &=& -3\left[\frac{P+P_B}{\rho}m\right]_0^R +  3\frac{\Gamma_1-1}{\Gamma_1}\int_0^R\frac{dP_B}{\rho}m \\
& +& \displaystyle{3\frac{\Gamma-1}{\Gamma}\int_0^R\,\frac{dP}{\rho}m}.
 \end{eqnarray}
The second term on the RHS of equation~(\ref{W6}) can be further recast to
\begin{center}
\begin{equation}
\begin{aligned}
{}& 3\frac{\Gamma_1 - 1}{\Gamma_1}\left(\int_0^RGm^2\,d\left(\frac{1}{r}\right) - \int_0^R\,\frac{dP}{\rho}m\right)\\
& =  3\frac{\Gamma_1 - 1}{\Gamma_1}\left(\frac{GM^2}{R} + 2W - \int_0^R\,\frac{dP}{\rho}m\right)
  \end{aligned}
 \end{equation}
\end{center}
and, using equations~(\ref{imp2}) and~(\ref{W6}), we obtain
\begin{center}
\begin{equation}
\begin{aligned}
W &= -3\left[\frac{P+P_B}{\rho}m\right]_0^R +  3\frac{\Gamma_1 - 1}{\Gamma_1}\left(\frac{GM^2}{R} + 2W\right)\\
&+ 3\left(\frac{\Gamma - 1}{\Gamma} - \frac{\Gamma_1 - 1}{\Gamma_1}\right) \int_0^R\,\frac{dP}{\rho}m.
  \end{aligned}
 \end{equation}
\end{center}

The first term vanishes because both forms of pressure are negligibly
small at the surface and mass vanishes at the centre.  Solving with the
expression for $P$ and using the same prescription as before we
obtain
\begin{center}
  \begin{equation}
    W = -\displaystyle{\frac{3(\Gamma_1-1)}{5\Gamma_1 -
        6}\frac{GM^2}{R}} + \displaystyle{\frac{\Gamma -
        \Gamma_1}{5\Gamma_1 - 6}\frac{3\Pi}{\Gamma - 1}},
 \end{equation}
\end{center}
assuming that $\rho$ is negligibly small at $r=R$, the surface of the
star, compared to the centre (or its average).  While computing
$\mu$ in this case, we integrate equation~(\ref{mu}) by simply
taking the average of $B$.  Although the integration is over $r$
(or $V$), we do not know a priori how $\rho$ or $B$ varies with $r$ in
this case.  So the integral in equation (\ref{mu}) simply gives
us $\Phi_M^2/6R$.  Thence, from equations~(\ref{virial1}),
(\ref{virial2}) and~(\ref{virial3}), we obtain


\begin{center}

\begin{equation}
\begin{aligned}
{}& -\displaystyle{\frac{3(\Gamma_1-1)}{5\Gamma_1 - 6}\frac{GM^2}{R}}
+ \displaystyle{\Big(1 + \frac{\Gamma - \Gamma_1}{(5\Gamma_1 -
    6)(\Gamma - 1)}\Big)\frac{3^\Gamma KM^\Gamma}{(4\pi
    R^3)^{\Gamma-1}}}\\ & + \frac{1}{6}\frac{\Phi_M^2}{R}= 0,
  \end{aligned}
 \end{equation}

\end{center}
and consequently

\begin{center}
\begin{equation}
\label{alpha_pl}
\alpha =\displaystyle{\frac{3(\Gamma_1-1)}{5\Gamma_1 - 6}}, 
\end{equation}

\begin{equation}
\label{beta_pl}
\beta =\displaystyle{\Big(1 + \frac{\Gamma - \Gamma_1}{(5\Gamma_1 - 6)(\Gamma - 1)}\Big)\frac{3^{\Gamma}K}{{(4\pi)}^{\Gamma-1}}}, 
\end{equation}
\begin{equation}
\label{gama_pl}
\gamma =\displaystyle{\frac{1}{6}}.
 \end{equation}
\end{center}
An important outcome here is that $\alpha$ is related to the scaling
of $B$ with $\rho$.  This is indeed expected from the
magnetohydrostatic balance equation~(\ref{mhd_eq}).  In other words, it
could be expected from equation~(\ref{mhd_eq}) itself that the presence
of magnetic pressure allows either a more massive or smaller star.  For
$\Gamma = \Gamma_1$ the result reduces to that of the nonmagnetic case
with a redefined $K$.



\subsection{Variation of $K$}
Equations~(\ref{beta_flfr}) and~(\ref{beta_pl}) derived above contain
$K$ which changes depending on the strength of the magnetic field.
For a weak magnetic field $B \lesssim 10^{14}\,$G, $\Gamma=4/3$
\citep{Subramanian2015}.  Thus, for this case we use Chandrasekhar's
\citep{Chandrasekhar1935}
\begin{center}
\begin{equation} 
K =
\frac{1}{8}\left(\frac{3}{\pi}\right)^\frac{1}{3}\left(\frac{hc}{(\mu_{\rm e}
  m_{\rm p})^\frac{1}{3}}\right),
\end{equation}
\end{center}
where $h$ is Planck's constant, $c$ is the speed of light, $\mu_{\rm e}
\approx 2$ is the mean molecular weight per electron and $m_{\rm p}$ is the
mass of proton.

The strong field $B \gtrsim 10^{16}\,$G case corresponds to $\Gamma
\approx 2$, because of Landau quantization.  For this case we
define $K$ as
\begin{center}
  \begin{equation}
    K = \displaystyle{\frac{m_{\rm e}c^2}{2Q\mu_{\rm e}m_{\rm p}}}
\end{equation}
\end{center}
\citep{Das2013}, where $m_{\rm e}$ is the mass of electron and $Q$ is given by
\begin{center}
  \begin{equation}
    Q = \displaystyle{\frac{\mu_{\rm e}m_{\rm p}B_{\rm D} }{2\pi^2\lambdabar_{\rm e}^3}},
\end{equation}
\end{center}                        
where $\lambdabar_{\rm e} = \frac{h}{2\pi m_{\rm e} c}$ is the reduced Compton
wavelength of electron and $B_{\rm D} = B/B_{\rm c}$ is the dimensionless magnetic
field, with $B_{\rm c} = 4.414 \times 10^{13}\,$G.

\section{Results}
\label{secresults}
We divide our findings into two classes, the \textit{Flux
  Conservation} model and the \textit{Power Law} model, mainly for
strongly magnetized (with $\overline{B}\approx 10^{16}\,$G) and weakly
magnetized (with $\overline{B}\lesssim 10^{14}\,$G) stars.  Ignoring
the thermal energy contribution for the time being, from
equation~(\ref{virial2}), regardless of the model or the strength of
the magnetic field, we have
\begin{center}
\begin{equation}
\label{rapprox}
R \approx \displaystyle{\left(\frac{\alpha GM^2 }{\gamma B^2}\right)^\frac{1}{4}}.
\end{equation}
\end{center}
Note that there is an inverse relation between radius and magnetic
field, while $\Gamma$ and $\Gamma_1$ do not increase much with
increasing $B$ or $\overline{B}$.  So an increase in the magnetic
field corresponds to a larger magnetic flux and this leads to a
contraction of the star in order to maintain virial equilibrium.  For
a B-WD of mass $2\,M_\odot$ and radius $1000\,$km, the maximum
$\overline{B}\approx 4\times 10^{13}\,$G for $\Gamma=1.8$ and $n=5$
according to the {\it Flux Conservation} model.  For the same
mass and radius the maximum $\overline{B}\approx 2\times 10^{14}\,$G
for $\Gamma_1=2$ according to the {\it Power Law} model.

We can now explore various properties of B-WDs based on either of the
models for various parameters.  All
the figures that follow have been based on equation~(\ref{virial3})
or~(\ref{M1}) and include all components of the energy.  We begin with
the \textit{Flux Conservation} model and consider the variation of $R$
with $n$.  For the strong field case ideally $\Gamma = 2$.  However
this may not be followed strictly so we also include the case when
$\Gamma = 1.8$.  For the \textit{Power Law} model we consider the
variation of $R$ with varying $\Gamma_1$, similarly to the previous
model.  The results are explored to determine whether $2\,M_\odot$,
$2.5\,M_\odot$ and~$3\,M_\odot$ stars are possible for either of the
models, because magnetic field is generally known to allow the
super-Chandrasekhar mass stars \citep{Ostriker1968,Das2013,Subramanian2015}.
\begin{figure}
\includegraphics[width=0.5\textwidth]{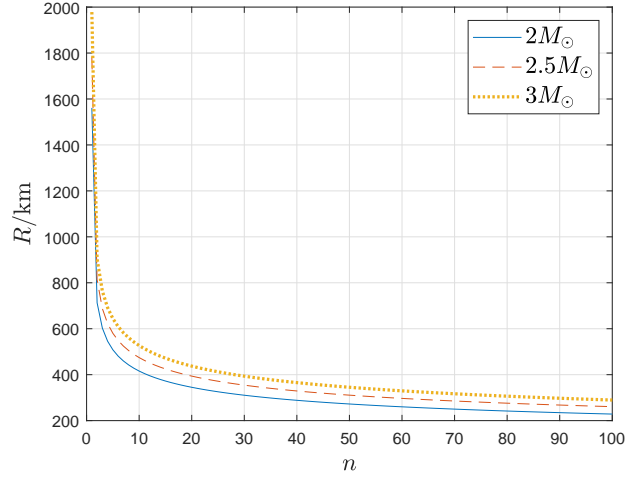}
\caption{Variation of the radius $R$ with $n$ for the \textit{Flux
    Conservation} model with $\Gamma=4/3$ and $\overline{B}=10^{14}\,$G.}
\label{fig:flfr14}
\end{figure}
\begin{figure}
\includegraphics[width=0.5\textwidth]{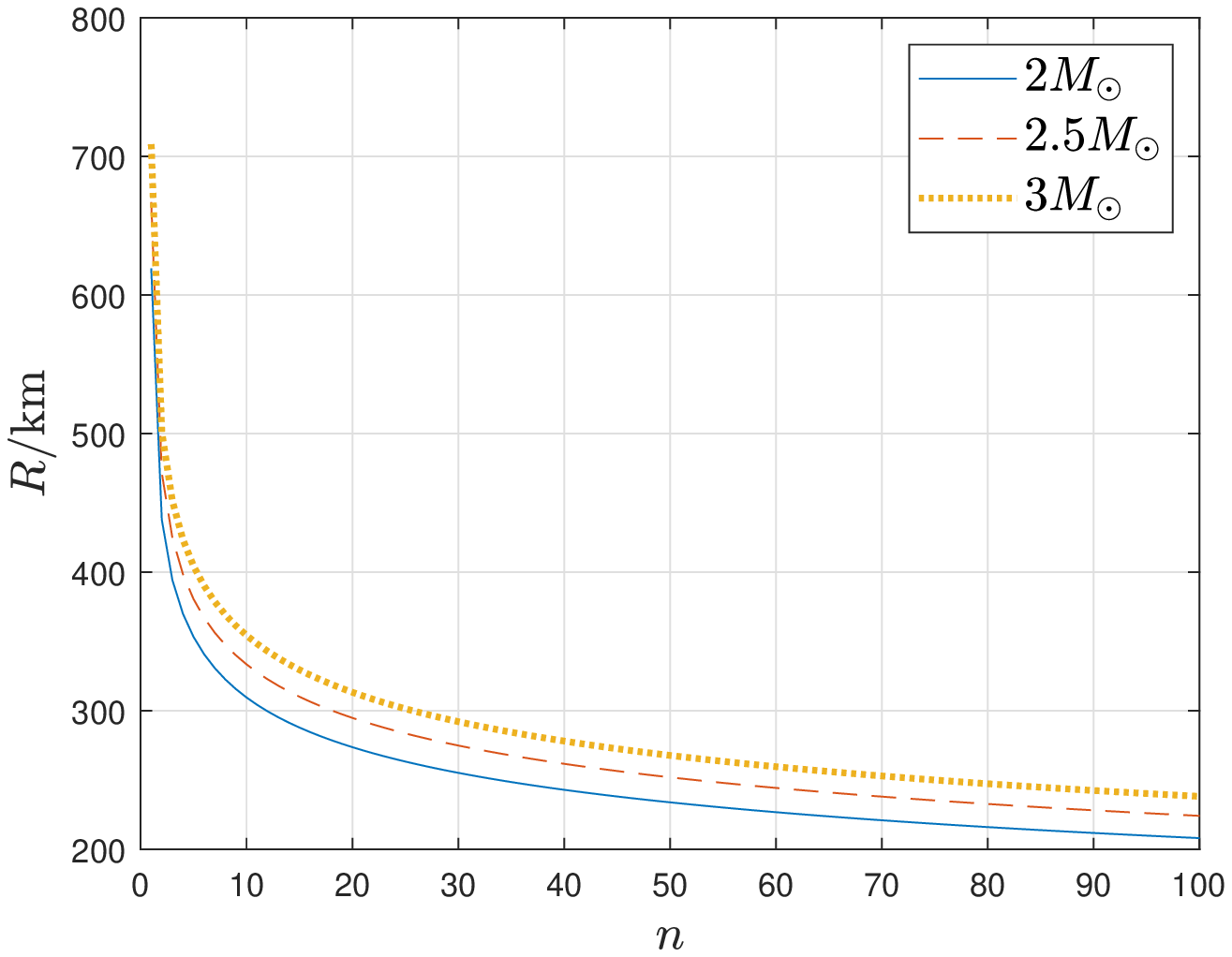}
\caption{Variation of the radius $R$ with $n$ for the \textit{Flux
    Conservation} model with $\Gamma=2$ and
  $\overline{B}=10^{16}\,$G.}
\label{fig:flfr162}
\end{figure}

\begin{figure}
\includegraphics[width=0.5\textwidth]{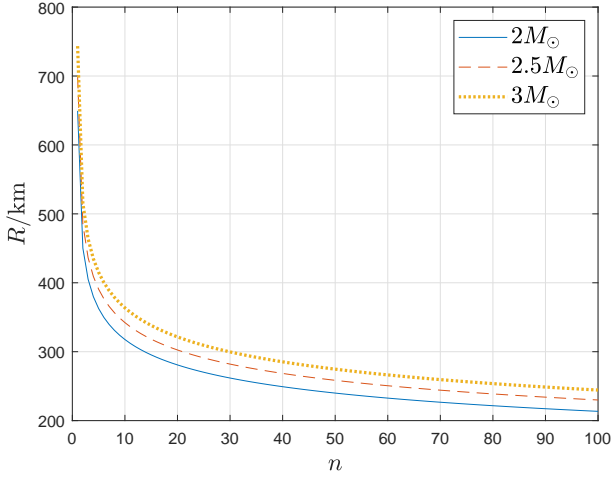}
\caption{Variation of the radius $R$ with $n$ for the \textit{Flux
    Conservation} model with $\Gamma=1.8$ and
  $\overline{B}=10^{16}\,$G.}
\label{fig:flfr1618}
\end{figure}

Figs~\ref{fig:flfr14}, \ref{fig:flfr162} and~\ref{fig:flfr1618} show
that, for a given mass, the radius decreases with increasing $n$.
This can be understood from equations~(\ref{alpha_flfr}),
(\ref{beta_flfr}) and~(\ref{gama_flfr}), where an increase in $n$
increases $\gamma$ but leaves $\alpha$ and~$\beta$ unchanged.  A
larger $\gamma$ results in a larger contribution of magnetic energy
and so, in order to maintain virial equilibrium, there must be an
increase in the star's potential energy, which results from a
contraction of the star.  Physically, this trend can be understood by
equation~(\ref{P_B}), wherein a smaller core leads to an overall
increase in total magnetic pressure, which is balanced by a larger
inward gravitational potential energy for a given mass.  So a smaller
core leads to a smaller star.
\begin{figure}
\includegraphics[width=0.5\textwidth]{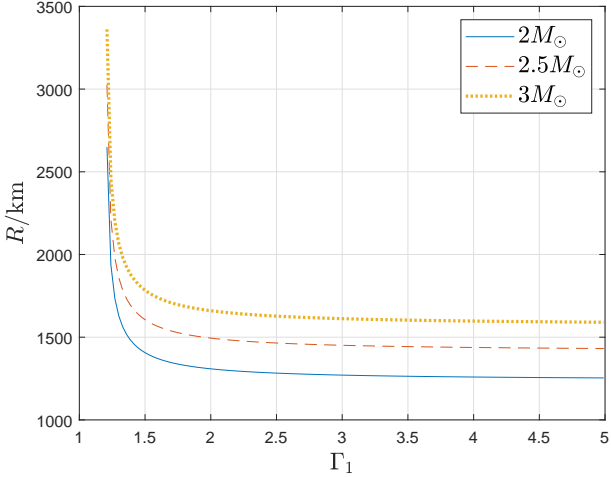}
\caption{Variation of the radius $R$ with $\Gamma_1$ for the
  \textit{Power Law} model with $\Gamma=4/3$ and
  $\overline{B}=10^{14}\,$G.}
\label{pl14}
\end{figure}
\begin{figure}
\includegraphics[width=0.5\textwidth]{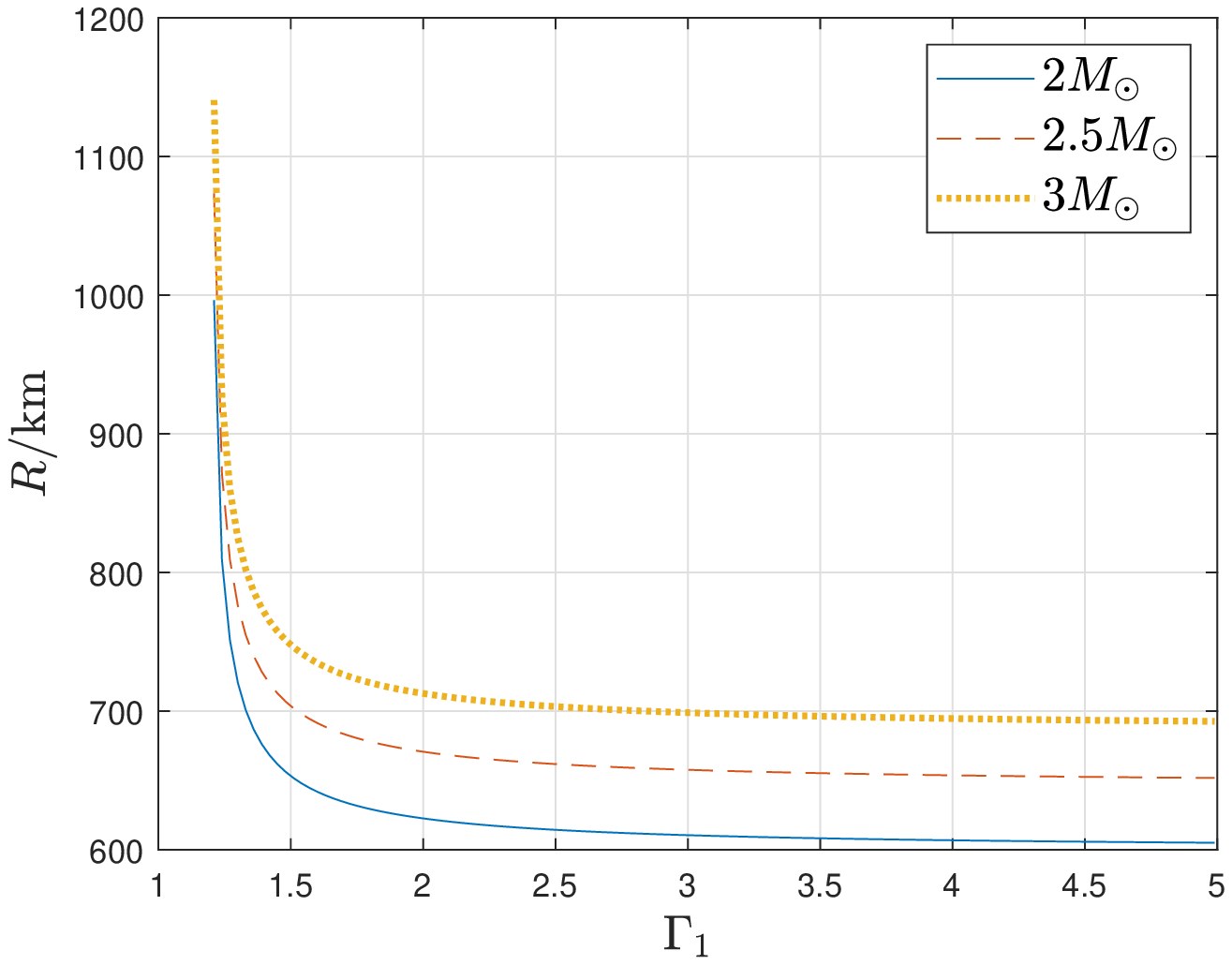}
\caption{Variation of the radius $R$ with $\Gamma_1$ for the
  \textit{Power Law} model with $\Gamma=2$ and
  $\overline{B}=10^{16}\,$G.}
\label{pl162}
\end{figure}
\begin{figure}
\includegraphics[width=0.5\textwidth]{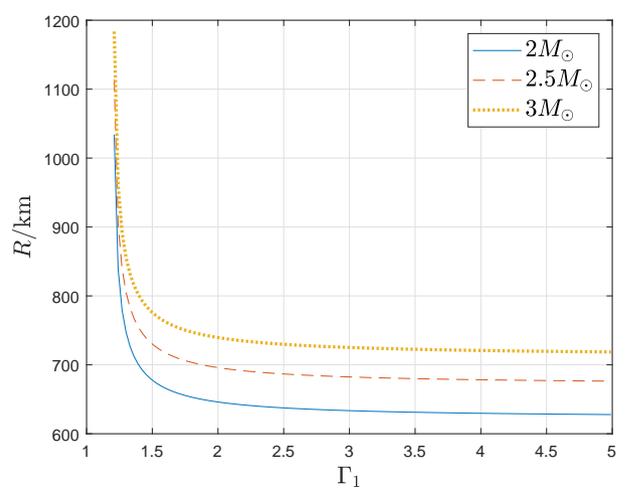}
\caption{Variation of the radius $R$ with $\Gamma_1$ for the
  \textit{Power Law} model with $\Gamma=1.8$ and
  $\overline{B}=10^{16}\,$G.}
\label{pl1618}
\end{figure}

Figs~\ref{pl14}, \ref{pl162} and~\ref{pl1618} have a region $\Gamma_1
\gtrsim 1.8$, where the radius tends to become independent of
$\Gamma_1$.  This can be understood from equation~(\ref{alpha_pl}),
where $\alpha$ is proportional to $(\Gamma_1 - 1)/(5\Gamma_1 - 6)$,
which becomes approximately a constant for $\Gamma_1 \gtrsim 1.8$.
This tells us that the power law dependence of $P_B$ on $\Gamma_1$ is
restricted to $\Gamma_1\lesssim 2$.  Also, $(\Gamma_1 - 1)/(5\Gamma_1
- 6)$ diverges for $\Gamma_1 = 1.2$ and becomes negative for
$1<\Gamma_1<1.2$.  This leads to an overall positive gravitational
potential energy, which would unbind the star.  So any physical
result must correspond to $\Gamma_1>1.2$.

\begin{figure}
\includegraphics[width=0.5\textwidth]{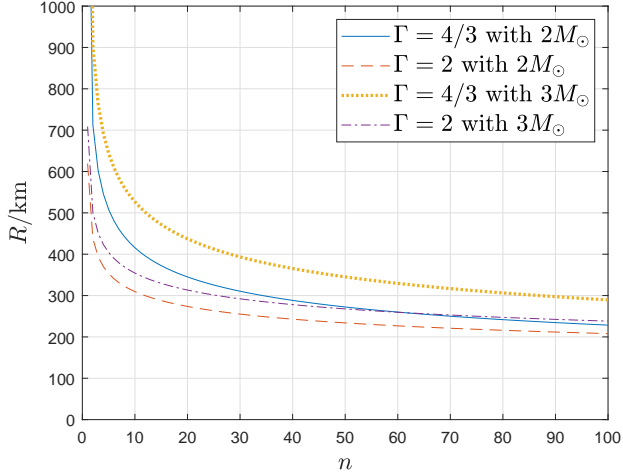}
\caption{Variation of the radius $R$ with $n$ for the \textit{Flux
    Conservation} model with various $\Gamma$ and total masses.  In each
  case $\Gamma = 4/3$ corresponds to $\overline{B}=10^{14}\,$G and
  $\Gamma=2$ corresponds to $\overline{B}=10^{16}\,$G.}
\label{crossflfr}
\end{figure}

\begin{figure}
\includegraphics[width=0.5\textwidth]{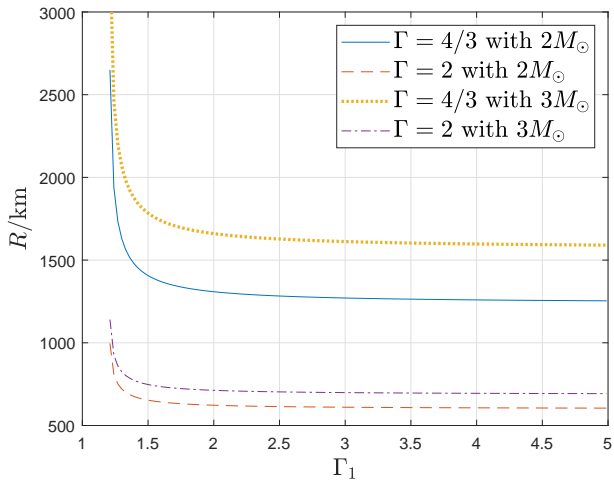}
\caption{Variation of the radius $R$ with $\Gamma_1$ for the \textit{Power
    Law} model with various $\Gamma$ and total masses.  In each case,
  $\Gamma = 4/3$ corresponds to $\overline{B}=10^{14}\,$G and
  $\Gamma=2$ corresponds to $\overline{B}=10^{16}\,$G.}
\label{crosspl}
\end{figure}

Furthermore, Figs~\ref{fig:flfr14}, \ref{fig:flfr162}
and~\ref{fig:flfr1618} with~\ref{pl14}, \ref{pl162} and~\ref{pl1618}
show that, for a given mass, with increasing $\Gamma$ there is a
decrease in the star's maximum attainable radius.  This can be
understood by solving equation~(\ref{virial3}) for various $\Gamma$.
However, decreasing $\Gamma$ corresponds to a significant decrease in
the magnetic field, by two orders of magnitude when $\Gamma$ falls from~2
to~4/3, and this increases $R$.  So we expect that $R$ for
$\Gamma = 4/3$ and a $2\,M_\odot$ star (curve A) is greater than $R$ for
$\Gamma = 2$ and a $3\,M_\odot$ star (curve B).  Figs~\ref{crossflfr}
and~\ref{crosspl} illustrate this for the \textit{Flux Conservation}
model and the \textit{Power Law} model respectively for $2\,M_\odot$
and $3\,M_\odot$ stars.  While \textit{curve A} is always above
\textit{curve B} throughout for the \textit{Power Law} model in
Fig.~\ref{crosspl}, indicating larger radii for the former, the
\textit{Flux Conservation} model in Fig.~\ref{crossflfr} shows an
intersection of the two curves.  This is due to the fact that there is
a stronger $n$ dependence in $\gamma$ for \textit{curve A} than
for \textit{curve B} (equations~\ref{gama_flfr} and \ref{rapprox}) and so the
magnetic flux dominates at higher $n$ for \textit{curve A}, thereby
driving it below \textit{curve B}, and hence decreasing $R$, above a
certain $n$.  Physically this may be thought of as the effect of high
magnetic flux density supporting gravity.

However equation~(\ref{virial3}), and consequently
equation~(\ref{rapprox}), cannot be used to calculate the radius of white
dwarfs with small magnetic field ($B\lesssim 10^{11}\,$G) unless the
magnetic flux is fixed.  For a fixed magnetic flux, decreasing field
increases the radius, the information pertaining to which is missing
in equation~(\ref{rapprox}).  More precisely, it is the $B^{-1/2}$
dependence of radius on magnetic field in equation~(\ref{rapprox})
which increases the radius extremely for small magnetic fields and
this is unphysical.  This is mainly because, at a lower $B$, the
contribution from thermal energy cannot be neglected compared with the
magnetic energy and so equation~(\ref{rapprox}) becomes invalid.


\section{Conclusions}
\label{secconclusions}
We have demonstrated the power of the highly magnetized virial theorem
to make broad statements about highly magnetized stars, particularly
white dwarfs.  The virial theorem is generally applicable to dynamical
and thermodynamic systems and can be formulated to address a
plethora of other systems, including relativistic systems and stars
with magnetic fields or rotation.  In the presence of additional
effects, the magnetic field in our case, application of the calculus
of variations to the theorem can provide information about dynamical
behaviour because it represents a structural relationship that the
system must follow.  However, we emphasize that the virial theorem is
an integral theorem that generally relates scalar quantities, three
different energy contributions in this case, rather than vectors.
Usually this reduction in complexity results in an associated loss of
information and we do not obtain as complete a description of a
physical system as would be possible from a complete analysis of the
system \citep{collins1978}.  Nevertheless, we have presented simple
analytical models of the properties of highly magnetized white dwarfs,
wherein important properties are revealed merely by looking at the
contributions of gravitational, thermal and magnetic energies.  We
have shown how these various contributions to energy change with the
introduction of the strong magnetic field when compared to nonmagnetic
or weakly magnetic counterparts.  This leads us to understand the
overall properties of a systems, in our case white dwarfs.

More precisely, we have explored the application of the virial theorem
to recently proposed B-WDs.  These highly magnetic white dwarfs can
explain several observations, including peculiar over-luminous type~Ia
supernovae, some white dwarf pulsars, soft gamma-ray repeaters and
anomalous X-ray pulsars \citep{mukhopadhyay2016soft}, all of which
have otherwise been rather puzzling.  We have shown that incorporating
magnetic field and thence magnetic pressure in the virial theorem can
explain the existence of super-Chandrasekhar white dwarfs with radii
significantly smaller than those of non-magnetized white dwarfs.  We
have explored this with two inherently different models: the first
considers a core with constant magnetic field and a varying field in
the outer envelope that conserves the magnetic flux; and the second
models the magnetic pressure as varying with the matter density as a
power law throughout.  Flux conservation is able to explain the
magnetic field variation and so the magnetic pressure variation of a
B-WD as well as a non-magnetic white dwarf.  Our chosen boundary
conditions, which are otherwise considered realistically for
  white dwarfs, play an important role when we obtain the
coefficients $\alpha$, $\beta$ and $\gamma$ in the various energy
contributions.  Nevertheless, under certain assumptions, which include
a $B$ profile that varies with the same slope for all sizes of central
core, our results show that a star might retain its spherical shape in
spite of the presence of strong magnetic field if it is non-rotating.
Importantly while a field profile needs to be prescribed in
  order to obtain our results, the same results are obtained from
  apparently different model profiles.  The only common feature is how
  the field varies, in one model directly with the radial coordinate
  and in the other with the stellar density which also falls with
  radius.  This suggests that it is not really the model profile,
  but the effect of the magnetic field in general which reveals new physics in
  the magnetized virial theorem.

A more detailed and rigorous study of these magnetized objects would uncover 
more of their inconspicuous features.  Nevertheless, as preliminary global
estimates of strongly magnetized stellar properties, including how
strong a magnetic field could be maintained in a star, this modified
virial theorem serves as a very useful tool.

\section*{Acknowledgements}
AS thanks the Department of Physics, Indian Institute of Science,
Bangalore, for providing internships during the winters of 2017 and
2019, where most of the work was composed.  CAT thanks Churchill
College for his fellowship. The work is partly supported by 
a project of Department of Science and Technology (DST-SERB), India, with 
Grant No. DSTO/PPH/BMP/1946 (EMR/2017/001226).  Thanks are also
due to the referee for encouraging comment and suggestion to improve
the presentation of the work.

\section*{Data availability}

No new data were generated or analysed in support of this research.
Any numerical codes and related data generated during the work will
be made available whenever required by the readers.


\begin{thebibliography}{}
\makeatletter
\relax
\def\mn@urlcharsother{\let\do\@makeother \do\$\do\&\do\#\do\^\do\_\do\%\do\~}
\def\mn@doi{\begingroup\mn@urlcharsother \@ifnextchar [ {\mn@doi@}
  {\mn@doi@[]}}
\def\mn@doi@[#1]#2{\def\@tempa{#1}\ifx\@tempa\@empty \href
  {http://dx.doi.org/#2} {doi:#2}\else \href {http://dx.doi.org/#2} {#1}\fi
  \endgroup}
\def\mn@eprint#1#2{\mn@eprint@#1:#2::\@nil}
\def\mn@eprint@arXiv#1{\href {http://arxiv.org/abs/#1} {{\tt arXiv:#1}}}
\def\mn@eprint@dblp#1{\href {http://dblp.uni-trier.de/rec/bibtex/#1.xml}
  {dblp:#1}}
\def\mn@eprint@#1:#2:#3:#4\@nil{\def\@tempa {#1}\def\@tempb {#2}\def\@tempc
  {#3}\ifx \@tempc \@empty \let \@tempc \@tempb \let \@tempb \@tempa \fi \ifx
  \@tempb \@empty \def\@tempb {arXiv}\fi \@ifundefined
  {mn@eprint@\@tempb}{\@tempb:\@tempc}{\expandafter \expandafter \csname
  mn@eprint@\@tempb\endcsname \expandafter{\@tempc}}}

\bibitem[\protect\citeauthoryear{Bandyopadhyay, Chakrabarty \& Pal}{Bandyopadhyay et~al.}{1997}]{debades}
Bandyopadhyay D., Chakrabarty S., Pal S., 1997, Phys. Rev. Lett.,
79, 2176

\bibitem[\protect\citeauthoryear{Banerjee, Shankar \& Singh}{Banerjee
  et~al.}{2017}]{Banerjee2017}Banerjee S., Shankar S., Singh T. P., 2017, J. Cosmol. Astropart. Phys.,
2017, 004

\bibitem[\protect\citeauthoryear{Belyaev, Ricci, {\v{S}}imkovic, Adam, Tater
  \& Truhl{\'{\i}}k}{Belyaev et~al.}{2015}]{Belyaev2015}
Belyaev V. B., Ricci P., {\v S}imkovic F., Adam J., Tater M., Truhl{\'\i}k E., 2015,
Nucl. Phys. A, 937, 17

\bibitem[\protect\citeauthoryear{Bera \& Bhattacharya}{Bera \&
  Bhattacharya}{2016}]{Bera2016}
Bera P.,  Bhattacharya D.,  2016, \mn@doi MNRAS, 465, 4026

\bibitem[\protect\citeauthoryear{Bertolami \& Mariji}{Bertolami \&
  Mariji}{2016}]{Bertolami2016}
Bertolami O.,  Mariji H.,  2016, Phys. Rev.~D, 93, 104046

\bibitem[\protect\citeauthoryear{Bhattacharya, Mukhopadhyay  \&
  Mukerjee}{Bhattacharya et~al.}{2018}]{Bhattacharya2018}
Bhattacharya M.,  Mukhopadhyay B.,   Mukerjee S.,  2018, \mn@doi MNRAS, 477, 2705

\bibitem[\protect\citeauthoryear{Carvalho, Arba{\~{n}}il, Marinho  \&
  Malheiro}{Carvalho et~al.}{2018}]{Carvalho2018}
Carvalho G.~A.,  Arba{\~{n}}il J. D.~V.,  Marinho R.~M.,   Malheiro M.,  2018, Eur.
Phys. J. C, 78, 411


\bibitem[\protect\citeauthoryear{Chandrasekhar}{Chandrasekhar}{1935}]{Chandrasekhar1935}
Chandrasekhar S.,  1935, MNRAS, 95, 207

\bibitem[\protect\citeauthoryear{Chatterjee, Fantina, Chamel, Novak  \&
  Oertel}{Chatterjee et~al.}{2017}]{Chatterjee2017}
Chatterjee D.,  Fantina A.~F.,  Chamel N.,  Novak J.,   Oertel M.,  2017,
MNRAS, 469, 95

\bibitem[\protect\citeauthoryear{Coelho, Marinho, Malheiro, Negreiros,
  C{\'{a}}ceres, Rueda  \& Ruffini}{Coelho et~al.}{2014}]{Coelho2014}
Coelho J.~G.,  Marinho R.~M.,  Malheiro M.,  Negreiros R.,  C{\'{a}}ceres
  D.~L.,  Rueda J.~A.,   Ruffini R.,  2014, ApJ, 794, 86

\bibitem[\protect\citeauthoryear{Collins}{Collins}{1978}]{collins1978}
  Collins G. W. II, 1978, Astron. Astrophys. Ser. Vol.~7, The Virial Theorem in
Stellar Astrophysics. Pachart Publishing House, Tucson, p.~143

\bibitem[\protect\citeauthoryear{Das \& Mukhopadhyay}{Das \&
  Mukhopadhyay}{2012}]{DAS2012}
  Das U.,  Mukhopadhyay B.,  2012, Int. J. Mod. Phys. D, 21, 1242001

\bibitem[\protect\citeauthoryear{Das \& Mukhopadhyay}{Das \&
  Mukhopadhyay}{2013}]{Das2013}
Das U.,  Mukhopadhyay B.,  2013, Phys. Rev. Lett., 110, 071102

\bibitem[\protect\citeauthoryear{Das \& Mukhopadhyay}{Das \&
  Mukhopadhyay}{2014a}]{Das2014}
Das U.,  Mukhopadhyay B.,  2014a, Mod. Phys. Lett. A, 29, 1450035

\bibitem[\protect\citeauthoryear{Das \& Mukhopadhyay}{Das \&
  Mukhopadhyay}{2014b}]{Das2014_1}
  Das U.,  Mukhopadhyay B.,  2014b, J. Cosmol. Astropart. Phys., 6, 050

\bibitem[\protect\citeauthoryear{Das \& Mukhopadhyay}{Das \&
  Mukhopadhyay}{2015}]{das15}
  Das U.,  Mukhopadhyay B.,  2015, J. Cosmol. Astropart. Phys., 05, 016

\bibitem[\protect\citeauthoryear{Dong, Zuo, Yin  \& Gu}{Dong
  et~al.}{2014}]{Dong2014}
Dong J.,  Zuo W.,  Yin P.,   Gu J.,  2014, Phys. Rev. Lett., 112, 039001

\bibitem[\protect\citeauthoryear{Eldridge \& Tout}{Eldridge \&
    Tout}{2019}]{Eldridge2019}
Eldridge J. J., Tout C. A., 2019, The Structure and Evolution of Stars. World
Scientific Publishing Europe, London
  
\bibitem[\protect\citeauthoryear{Federbush, Luo  \& Smoller}{Federbush
  et~al.}{2014}]{Federbush2014}
Federbush P.,  Luo T.,   Smoller J.,  2014, Archive for Rational
  Mechanics and Analysis, 215, 611

\bibitem[\protect\citeauthoryear{Fendt \& Dravins}{Fendt \& Dravins}
{2000}]{fendt}
Fendt C.,  Dravins D., 2000, Astron. Nachr., 321, 193

\bibitem[\protect\citeauthoryear{Franzon \& Schramm}{Franzon \&
  Schramm}{2015}]{Franzon2015}
Franzon B.,  Schramm S.,  2015, Phys. Rev. D, 92, 083006

\bibitem[\protect\citeauthoryear{Franzon \& Schramm}{Franzon \&
  Schramm}{2017}]{Franzon2017}
Franzon B.,  Schramm S.,  2017, MNRAS, 467, 4484

\bibitem[\protect\citeauthoryear{Gupta, Mukhopadhyay  \& Tout}{Gupta
  et~al.}{2020}]{Gupta2020}
Gupta A.,  Mukhopadhyay B.,   Tout C.~A.,  2020, MNRAS, 496, 894

\bibitem[\protect\citeauthoryear{Herrera \& Barreto}{Herrera \&
  Barreto}{2013}]{Herrera2013}
Herrera L.,  Barreto W.,  2013, Phys. Rev. D, 87, 087303

\bibitem[\protect\citeauthoryear{Howell et~al.,}{Howell
  et~al.}{2006}]{AndrewHowell2006}
Howell D.~A.,  et~al., 2006, Nature, 443, 308

\bibitem[\protect\citeauthoryear{Kalita \& Mukhopadhyay}{Kalita \&
  Mukhopadhyay}{2019}]{kalita2019continuous}
  Kalita S.,  Mukhopadhyay B.,  2019, MNRAS, 490, 2692

\bibitem[\protect\citeauthoryear{Kalita, Mukhopadhyay, Mondal  \& Bulik}{Kalita
  et~al.}{2020}]{Kalita2020}
Kalita S.,  Mukhopadhyay B.,  Mondal T.,   Bulik T.,  2020, ApJ, 896, 69

\bibitem[\protect\citeauthoryear{Kundu \& Mukhopadhyay}{Kundu \&
  Mukhopadhyay}{2012}]{KUNDU2012}
Kundu A.,  Mukhopadhyay B.,  2012, Mod. Phys. Lett. A, 27, 1250084

\bibitem[\protect\citeauthoryear{Liu, Zhang  \& Wen}{Liu
  et~al.}{2014}]{Liu2014}
Liu H.,  Zhang X.,   Wen D.,  2014, Phys. Rev. D, 89, 104043

\bibitem[\protect\citeauthoryear{Markey \& Tayler}{Markey \&
  Tayler}{1973}]{Markey1973}
Markey P.,  Tayler R.~J.,  1973, MNRAS, 163, 77

\bibitem[\protect\citeauthoryear{Moussa}{Moussa}{2017}]{Moussa2017}
Moussa M.,  2017, Annals of Physics, 385, 347

\bibitem[\protect\citeauthoryear{Mukhopadhyay \& Rao}{Mukhopadhyay \&
    Rao}{2016a}]{mukhopadhyay2016soft}
  Mukhopadhyay B., Rao A. R., 2016, J. Cosmol. Astropart. Phys., 5, 007

\bibitem[\protect\citeauthoryear{Mukhopadhyay et al.}{Mukhopadhyay
  et~al.}{2016}]{mukhopadhyay2016significantly}
Mukhopadhyay B., Das U., Rao A. R., Subramanian S., Bhattacharya M.,
Mukerjee S., Bhatia T. S., Sutradhar J., 2017, in Tremblay P.-E., G{\"a}nsicke
B., Marsh T., eds, Astronomical Society of the Pacific Conference Series
Vol. 509, 20th European White Dwarf Workshop. Astron. Soc. Pac., San
Francisco, p.~401

\bibitem[\protect\citeauthoryear{Mukhopadhyay, Rao \& Bhatia}{Mukhopadhyay
et~al.}{2017}]{mukhopadhyaySco}
  Mukhopadhyay B., Rao A. R., Bhatia, T. S., 2017, MNRAS, 472, 3564

\bibitem[\protect\citeauthoryear{Narayan \& Yi}{Narayan \&
  Yi}{1994}]{narayanyi}
Narayan R.,  Yi I., 1994, ApJ, 428, L13

\bibitem[\protect\citeauthoryear{Ostriker \& Hartwick}{Ostriker \&
  Hartwick}{1968}]{Ostriker1968}
Ostriker J.~P.,  Hartwick F. D.~A.,  1968, ApJ, 153, 797

\bibitem[\protect\citeauthoryear{Otoniel, Franzon, Carvalho, Malheiro, Schramm
  \& Weber}{Otoniel et~al.}{2019}]{Otoniel2019}
Otoniel E.,  Franzon B.,  Carvalho G.~A.,  Malheiro M.,  Schramm S.,   Weber
  F.,  2019, ApJ, 879, 46

\bibitem[\protect\citeauthoryear{Eslam Panah \& Liu}{Eslam Panah \&
  Liu}{2019}]{EslamPanah2019}
Eslam Panah B.~E.,  Liu H.,  2019, Phys. Rev. D, 99, 104074

\bibitem[\protect\citeauthoryear{Pili, Bucciantini \& Del Zanna}{Pili et~al.}{2014}]{pili}
Pili A. G., Bucciantini N., Del Zanna L., 2014, MNRAS, 439, 3541

\bibitem[\protect\citeauthoryear{Pons \& Vigan\'o}{Pons \& Vigan\'o}{2019}]{pons}
Pons J. A., Vigan\'o D., 2019, Liv. Rev. Com. Astrophys., 5, 3

\bibitem[\protect\citeauthoryear{Quentin \& Tout}{Quentin \&
  Tout}{2018}]{tout}
Quentin L.~G.,  Tout C.~A., 2018, MNRAS, 477, 2298

\bibitem[\protect\citeauthoryear{Roy, Mukhopadhyay, Lahiri  \& Basu}{Roy
  et~al.}{2019}]{Roy2019}
Roy S.~K.,  Mukhopadhyay S.,  Lahiri J.,   Basu D.,  2019, Phys. Rev. D, 100, 063008

\bibitem[\protect\citeauthoryear{Scalzo et~al.,}{Scalzo
  et~al.}{2010}]{Scalzo2010}
Scalzo R.~A., et~al., 2010, ApJ, 713, 1073

\bibitem[\protect\citeauthoryear{Shah \& Sebastian}{Shah \&
  Sebastian}{2017}]{Shah2017}
Shah H.,  Sebastian K.,  2017, ApJ, 843, 131

\bibitem[\protect\citeauthoryear{Shapiro \& Teukolsky}{Shapiro \&
    Teukolsky}{1983}]{Shapiro:1983du}
Shapiro S. L., Teukolsky S. A., 1983, Black Holes, White Dwarfs and
Neutron Stars: The Physics of Compact Objects. Wiley, New York

\bibitem[\protect\citeauthoryear{Sotani \& Tatsumi}{Sotani \&
  Tatsumi}{2017}]{Sotani2017}
Sotani H.,  Tatsumi T.,  2017, MNRAS, 467, 1249

\bibitem[\protect\citeauthoryear{Subramanian \& Mukhopadhyay}{Subramanian \&
  Mukhopadhyay}{2015}]{Subramanian2015}
Subramanian S.,  Mukhopadhyay B.,  2015, MNRAS, 454, 752

\bibitem[\protect\citeauthoryear{Tayler}{Tayler}{1973}]{Tayler1973}
Tayler R.~J.,  1973, MNRAS, 161, 365

\bibitem[\protect\citeauthoryear{Wickramasinghe, Tout  \&
  Ferrario}{Wickramasinghe et~al.}{2013}]{Wickramasinghe2013}
Wickramasinghe D.~T.,  Tout C.~A.,   Ferrario L.,  2013, MNRAS, 437, 675

\makeatother
\end{thebibliography}
\bsp	
\label{lastpage}
\end{document}